\documentclass[12pt,tightenlines,nofootinbib,aps,floatfix,pra,eqsecnum]{revtex4-1}
\usepackage{booktabs}
\usepackage{dcolumn}
\usepackage{units}
\usepackage{array}

\usepackage[utf8]{inputenc}
\usepackage{amsfonts}
\usepackage{amsmath,amssymb}
\usepackage{graphicx}
\usepackage{mathtools}
\usepackage[section]{placeins}
\usepackage{float}
\usepackage[font=footnotesize,labelfont=bf]{caption}
\usepackage{alphalph}
\usepackage{natbib}
\usepackage{textcomp}
\usepackage{amsmath}
\usepackage{amssymb}
\usepackage{multirow}
\usepackage{braket}
\usepackage{slashed}
\usepackage{color}

\usepackage{float}
\usepackage{subfig}
\usepackage{multirow}
\usepackage{hhline}



\DeclareMathOperator{\Tr}{Tr}

\begin{document}

%\title{Fermion-Dimer Scattering using Impurity Lattice Monte Carlo and the Adiabatic Projection Method\\  }

\title{Fermion-dimer scattering using an impurity lattice Monte Carlo
approach and the adiabatic projection method\\  }

\author{Serdar Elhatisari}
\email{selhati@ncsu.edu}
\affiliation{Department of Physics, North Carolina State University,
Raleigh, North Carolina 27695, USA}

\author{Dean Lee}
\email{dean_lee@ncsu.edu}
\affiliation{Department of Physics, North Carolina State University,
Raleigh, North Carolina 27695, USA}

\date{\today}

\begin{abstract}
We present lattice Monte Carlo calculations of fermion-dimer scattering in the limit of zero-range interactions using the adiabatic projection method.  The adiabatic projection method uses a set of initial cluster states and Euclidean time projection to  give a systematically improvable description of the low-lying scattering   cluster states in a finite volume.  We use L{\"u}scher's finite-volume relations to determine the $s$-wave, $p$-wave, and $d$-wave phase shifts.  For comparison, we also compute exact lattice results using Lanczos iteration and continuum results using the Skorniakov-Ter-Martirosian equation.  For our Monte Carlo calculations we use a new lattice algorithm called impurity lattice Monte Carlo.  This algorithm can be viewed as a hybrid technique which incorporates  elements of both worldline and auxiliary-field Monte Carlo simulations.   \end{abstract}

\maketitle

\section{Introduction}
\label{sec:introduction}
The adiabatic projection method is a general  framework for calculating scattering and reactions  on the lattice.  The method constructs a low-energy effective theory for clusters which becomes exact in the limit of large Euclidean projection time. Previous studies of this method \cite{Rupak:2013aue,Pine:2013zja} have used exact sparse matrix methods.  In this work we demonstrate the first application using Monte Carlo simulations.  As we will show, the adiabatic projection method significantly improves the  accurate calculation of finite-volume energy levels.  As we also will show, the
finite-volume energy levels must be calculated with considerable accuracy to determine the scattering phase shifts using L\"uscher's method.  We give a short summary of L\"uscher's method later in our discussion.

The goal of this analysis is to benchmark the use of lattice Monte Carlo simulations with the adiabatic projection method.  The example we consider in detail is fermion-dimer scattering for two-component fermions and zero-range interactions.  Our  calculation also corresponds to neutron-deuteron scattering in the spin-quartet channel at leading order in pionless effective field theory. In our interacting system there are two components for the fermions.  We call the two components up and down spins, $\uparrow$ and $\downarrow$.   The bound dimer state is composed of one $\uparrow$ and one $\downarrow$, and our fermion-dimer system consists of two $\uparrow$  and one $\downarrow$.  While $s$-wave scattering was considered previously \cite{PhysRevD.84.091503,PhysRevC.86.034003,A.Rokash2013.3386,Pine:2013zja}, we will present the first lattice calculations of $p$-wave and $d$-wave fermion-dimer scattering. 

As discussed in Ref.~\cite{Pine:2013zja}, the adiabatic projection method starts with a set of initial cluster states.  By clusters we mean either a single particle or a bound state of several particles.   In our analysis here we consider fermion-dimer elastic scattering where there are two clusters.  In Ref.~\cite{Pine:2013zja}, the initial fermion-dimer states were parametrized by the initial spatial separation between clusters, $\vec{R}$.  The initial cluster states can be written explicitly as
\begin{equation}
\lvert \vec{R} \rangle = \sum_{\vec{n}}b^{\dagger}_{\uparrow}(\vec{n})b^{\dagger}_{\downarrow}(\vec{n})b^{\dagger}_{\uparrow}(\vec{n}+\vec{R}) \lvert 0 \rangle,
\end{equation}
where the spatial volume is a periodic cubic box of length $L$ in lattice units.  The initial states are then projected using Euclidean time to form dressed cluster states,
 \begin{equation}
\lvert \vec{R} \rangle_t = e^{-\hat{H}t}\lvert \vec{R} \rangle. \end{equation}
The adiabatic method uses these dressed cluster states to calculate matrix elements of the Hamiltonian and other observables.  The result is a low-energy effective theory of interacting clusters which becomes systematically more accurate as the projection time $t$ is increased.
 An estimate of the residual error is derived in Ref.~\cite{Pine:2013zja}. 

For our calculations here we follow the same general process except that we build the initial cluster states in a different manner. Instead of working with the relative separation between clusters, we work with the relative momentum between the clusters.  We find that this change improves the efficiency of the Monte Carlo calculation by reducing the number of required initial states.  The new technique involves first constructing  a dimer state with momentum $\vec{p}$ using Euclidean time projection and then   multiplying by a creation operator for a second $\uparrow$ particle with momentum $-\vec{p}$.  For example, we can write the initial fermion-dimer state explicitly as
\begin{equation}
\lvert \vec{p} \rangle = \tilde{b}^{\dagger}_{\uparrow}(-\vec{p}) e^{-\hat{H}t'} \, \tilde{b}^{\dagger}_{\uparrow}(\vec{p})\tilde{b}^{\dagger}_{\downarrow}(\vec{0})\lvert 0 \rangle
\label{eqn:initialState-0001}\,.
\end{equation}
From these states we produce dressed cluster states by Euclidean time projection,
\begin{equation}
\lvert \vec{p} \rangle_t = e^{-\hat{H}t/2} \lvert \vec{p} \rangle
\label{eqn:ClusterState-0001}\,.
\end{equation}
We then proceed in the same manner as in Ref.~\cite{Pine:2013zja} and calculate the matrix elements of the Hamiltonian in the basis of the dressed cluster states.   

For our Monte Carlo simulations we introduce a new algorithm which we call the impurity lattice Monte Carlo 
algorithm.  Credit for developing this algorithm is to be shared with Ref.~\cite{Bour:2014}, where applications to impurities in many-body systems are being investigated using the same method.  It can be viewed as a hybrid algorithm in between worldline and auxiliary-field Monte Carlo simulations.  In worldline algorithms, the quantum amplitude is calculated by sampling particle worldlines in Euclidean spacetime.  In auxiliary-field Monte Carlo simulations, the interactions are recast as single particle interactions, and the  quantum amplitude is computed exactly for each auxiliary field configuration.  In impurity Monte Carlo, we handle the impurities using worldline Monte Carlo simulations while all other particles are treated using the auxiliary-field formalism.  Furthermore, the impurity worldlines themselves are acting as additional auxiliary fields felt by other particles in the system.  We have found that for our system of two $\uparrow$ and one $\downarrow$ particles, the impurity lattice Monte Carlo method is computationally superior to other methods such as the auxiliary-field Monte Carlo because of its speed and efficiency as well as control over sign oscillations. We will derive the formalism of impurity Monte Carlo simulations in detail in our discussion here. 

The organization of our paper is as follows.  We first start with the basic continuum and lattice formulations of our interacting system with zero-range two-component fermions.  We then take a short detour to derive the connection between normal-ordered transfer matrices and lattice Grassmann actions.  Using our dictionary between lattice Grassmann actions and quantum operators, we derive the transfer matrix induced by a given single impurity worldline.  We then describe the implementation of the adiabatic projection method and the details of our Monte Carlo simulations for computing finite-volume energy levels. 

To determine scattering phase shifts, we then discuss L\"uscher's finite-volume method.  As part of this discussion we discuss for the first time, the character of topological volume corrections for fermion-dimer scattering in the $p$-wave and $d$-wave channels.  By topological volume corrections, we are specifically referring to  momentum-dependent finite-volume corrections of the  dimer binding energy \cite{PhysRevD.84.091503,Davoudi:2011md}.  Previous studies looking at topological volume corrections had only considered $s$-wave scattering \cite{PhysRevD.84.091503,PhysRevC.86.034003,A.Rokash2013.3386,Pine:2013zja}.  The extension to higher partial waves is given in the appendix.  We then conclude with a comparison of Monte Carlo results as well as exact lattice calculations and continuum calculations.

\section{Lattice Hamiltonian}

We consider a three-body system of two-component fermions with equal mass, $m_{\uparrow}=m_{\downarrow}=m$.
We consider the  limit of large scattering length between the two components  where the interaction range of the fermions is taken to be negligible. We start with the free nonrelativistic Hamiltonian,
\begin{align}
 \hat{H}_{0} = \frac{1}{2m} \sum_{s=\uparrow,\downarrow}
 \int d^{3}\vec{r} \, \,  \vec{\nabla} b_{s}^{\dagger}(\vec{r}) \cdot \vec{\nabla} b_{s}(\vec{r}) \,,
 \label{eqn:free-Hamiltonian-001}
\end{align}
In the low-energy limit the interaction can be simplified as a delta-function interaction between  the two spin components, \begin{align}
 \hat{H} = \frac{1}{2m} \sum_{s=\uparrow,\downarrow}
 \int d^{3}\vec{r} \, \,  \vec{\nabla} b_{s}^{\dagger}(\vec{r}) \cdot \vec{\nabla} b_{s}(\vec{r})
 +
 C_{0} \int d^{3}\vec{r}  \, \,
\hat{\rho}_{\uparrow}(\vec{r}) \,  \hat{\rho}_{\downarrow}(\vec{r}) \,,
 \label{eqn:Hamiltonian-001}
 \end{align}
 where $\hat{\rho}_{\uparrow,\downarrow}(\vec{r})$ are density operators,
 \begin{align}
\hat{\rho}_{\uparrow}(\vec{r}) =  b_{\uparrow}^{\dagger}(\vec{r}) b_{\uparrow}(\vec{r}) \,,
\\
\hat{\rho}_{\downarrow}(\vec{r}) =  b_{\downarrow}^{\dagger}(\vec{r}) b_{\downarrow}(\vec{r})\,.
 \label{eqn:Latt-densty-op-001}
 \end{align}
The ultraviolet physics of this zero-range interaction must be regulated in some manner.  In our case the lattice provides the needed regularization. We denote the spatial lattice spacing as $a$ and the temporal lattice spacing as $a_{t}$. We will write all quantities in lattice units, which are physical units multiplied by the corresponding power of $a$ to render the combination dimensionless. We define the free nonrelativistic lattice Hamiltonian as
\begin{align}
\hat{H}_{0} =\hat{H}^{\uparrow}_{0} + \hat{H}^{\downarrow}_{0}\,,
 \label{eqn:lattice-Hamiltonian-001}
\end{align}
where
\begin{align}
\hat{H}^s_0=\frac{1}{2m} \sum_{l=1}^{3}
 &
 \sum_{\vec{n}}
 \left[
 2b_{s}^{\dagger}(\vec{n})b_{s}(\vec{n})
 -b_{s}^{\dagger}(\vec{n})b_{s}(\vec{n}+\hat{l})
 -b_{s}^{\dagger}(\vec{n})b_{s}(\vec{n}-\hat{l})
 \right] \,,
\end{align}
and the contact interaction potential is
\begin{align}
\hat{V} = C_{0}
&\sum_{\vec{n}}
\hat{\rho}_{\uparrow}(\vec{n}) \,
\hat{\rho}_{\downarrow}(\vec{n})\,.
 \label{eqn:lattice-Hamiltonian-005}
\end{align}
Here $\hat{l}$ denotes a lattice unit vector in one of the spatial directions, $\hat{l}=\hat{1},\hat{2},\hat{3}$.  The unknown interaction coefficient $C_0$ is tuned to reproduce the desired binding energy of the dimer at infinite volume.

 \section{Lattice path integrals and transfer matrices}
\label{sec:LatticePIandTM}

In this section we introduce the transfer matrix formalism that we use  in both the Monte-Carlo simulations and exact lattice calculations. We use the same action as in Refs.~\cite{Lee:2008xsa,Lee:2008fa}.  We start by defining the lattice action in terms of Grassmann variables.  We then we give the exact connection between the Grassmann path integral and normal-ordered transfer matrices. Let $\theta$ and $\theta^{*}$ be anticommuting Grassmann variables. Our  lattice action can be decomposed into three parts,  
\begin{align}
  S[\theta,\theta^{*}]
  =\sum_{n_t} \left\{S_{t}[\theta,\theta^{*},n_{t}]+S_{H_0}[\theta,\theta^{*},n_t]
  +S_{V}[\theta,\theta^{*},n_{t}]\right\} \,,
 \label{eqn:Action-003}
\end{align}
 where $S_{t}$ and $S_{H_0}$ contain temporal hopping and spatial hopping terms of the free lattice action respectively,  
 \begin{align}
 S_{t}[\theta,\theta^{*},n_{t}] = \sum_{s=\uparrow,\downarrow} &\sum_{\vec{n}}
 \left[ \theta_{s}^{*}(n_{t}+\hat{0},\vec{n}) \, \theta_{s}(n_{t},\vec{n})-\theta_{s}^{*}(n_{t},\vec{n}) \, \theta_{s}(n_{t},\vec{n})
   \right]\,,
 \label{eqn:Lattice-Action-temporal-001}
 \end{align}
  \begin{align}
 S_{H_0}[\theta,\theta^{*},n_{t}] =\frac{\alpha_{t}}{2m} \sum_{s=\uparrow,\downarrow} &\sum_{_{}\vec{n}}
 \sum_{l=1}^{3}
 \, \theta_{s}^{*}(n_{t},\vec{n}) \,
  \left[2 \, \theta_{s}(n_{t},\vec{n})- \theta_{s}(n_{t},\vec{n}+\hat{l})-
  \theta_{s}(n_{t},\vec{n}-\hat{l}) \right] \,.
 \label{eqn:Lattice-Action-spatial-001}
 \end{align}
 Here $\hat{0}$ denotes the lattice unit vector in the forward temporal direction and $\alpha_{t}$ is the dimensionless ratio of the temporal lattice spacing to the spatial lattice spacing, $a_{t}/a$.
 We can also write $S_{H_0}[\theta,\theta^{*},n_{t}]$ as \begin{equation}
S_{H_0}[\theta,\theta^{*},n_{t}] =\alpha_{t}H^{\uparrow}_0[\theta_{\uparrow},\theta^*_{\uparrow},n_t]+\alpha_{t}H^{\downarrow}_0[\theta_{\downarrow},\theta^*_{\downarrow},n_t],
\end{equation}
where
\begin{equation}
H_0^s[\theta_{s},\theta^*_{s},n_{t}] = \frac{1}{2m}\sum_{_{}\vec{n}}
 \sum_{l=1}^{3}
 \, \theta_{s}^{*}(n_{t},\vec{n}) \,
  \left[2 \, \theta_{s}(n_{t},\vec{n})- \theta_{s}(n_{t},\vec{n}+\hat{l})-
  \theta_{s}(n_{t},\vec{n}-\hat{l}) \right].
\end{equation}The interaction term has the form,
   \begin{align}
 S_{V}[\theta,\theta^{*},n_{t}]
 =
  \alpha_{t} C_{0} \,  \sum_{_{}\vec{n}} \,  \theta_{\uparrow}^{*}(n_{t},\vec{n}) \, \theta_{\uparrow}(n_{t},\vec{n})
   \, \theta_{\downarrow}^{*}(n_{t},\vec{n}) \, \theta_{\downarrow}(n_{t},\vec{n})\,.
 \label{eqn:Lattice-Action-001}
 \end{align}
 We take our system to reside in a cubic box of length $L$ units in the spatial directions and $L_t$ in the temporal direction.  Our Grassmann variables are chosen to be periodic along the spatial directions and antiperiodic in the temporal direction.
The antiperiodic boundary conditions in time are necessary for the trace formula to come later in Eq.~(\ref{eqn:G-path-T-matrix-001}).  The Grassmann path integral has the form,
\begin{align}
 \mathcal{Z} = \int
 \left[\prod_{n_{t},\vec{n},s=\uparrow,\downarrow}
 d\theta_{s}(n_{t},\vec{n})d\theta_{s}^{*}(n_{t},\vec{n})\right]  \, \, e^{- S[\theta,\theta^{*}]} \,.
 \label{eqn:Grassmann-path-int-001}
 \end{align}

While the Grassmann formalism is convenient for deriving the lattice Feynman rules, the transfer matrix formalism is more  convenient for numerical calculations. To make the connection between the two formulations, using the materials given in Appendix~\ref{sec:append-operator-integral-formalism} we write the following exact relation between the Grassmann path integral formula and the transfer matrix formalism. For any function $f$,
 \begin{align}
 \Tr  & \Big\{ \, : \,  f_{L_{t}-1}[a_{s} (\vec{n}),a^{\dagger}_{s^{\prime}} (\vec{n^{\prime}})] \, : \,
 \cdots \, : \, f_{0}[a_{s} (\vec{n}),a^{\dagger}_{s^{\prime}} (\vec{n^{\prime}})] \, : \, \Big\}
 \nonumber\\
 &
 = \int
 \left[\prod_{n_{t},\vec{n},s=\uparrow,\downarrow}
 d\theta_{s}(n_{t},\vec{n})d\theta_{s}^{*}(n_{t},\vec{n})\right]  \, e^{-\sum_{n_t} S_{t}[\theta,\theta^{*},n_{t}]}
 \prod_{n_{t} = 0}^{L_{t}-1}
 f_{n_{t}}[\theta_{s} (n_{t},\vec{n}),\theta^{*}_{s^{\prime}} (n_{t},\vec{n}^{\prime})] \,,
 \label{eqn:G-path-T-matrix-001}
 \end{align}
 where the symbol :\,: signifies normal ordering. Normal ordering rearranges all operators so that all annihilation operators are moved to the right and creation operators are moved to the left with the appropriate number of anticommutation minus signs.  Then the desired transfer matrix formulation of the path integral is
 \begin{align}
 \mathcal{Z} = \Tr \,  \hat{M}^{L_{t}} \,,
 \label{eqn:transfer-matrix-0100}
 \end{align}
 where $\hat{M}$ is the normal-ordered transfer matrix operator,
 \begin{align}
 \hat{ M} = \,  : \, \exp
 \left[
 -\alpha_{t} \hat{H}_{0}-\alpha_{t} C_{0}
 \sum_{\vec{n}}
\hat{\rho}_{\uparrow}(\vec{n})
 \hat{\rho}_{\downarrow}(\vec{n})  \right]
 \, : \,.
 \label{eqn:transfer-matrix-0200}
 \end{align}
Here $\hat{H}_{0}$ is the free lattice Hamiltonian given in Eq.~(\ref{eqn:lattice-Hamiltonian-001}).

\section{Impurity Lattice Monte Carlo:  Single IMpurity}
\label{sec:ImpurityLMC}

In this section we derive the formalism for impurity lattice Monte Carlo for a single impurity.  In impurity Monte Carlo the impurities are treated differently from other particles.
The assumption is that there are only a small number of impurities and these can be sampled using worldline Monte Carlo without strong fermion sign oscillation problems from antisymmetrization.  In our case there is exactly one $\downarrow$  particle, and we treat this as a single impurity for our system.

Let us consider the occupation number basis,
\begin{align}
\Ket{\chi^\uparrow_{n_{t}},
\chi^{\downarrow}_{n_{t}}}
=
\prod_{\vec{n}}
\left\{
\left[
b_{\uparrow}^{\dagger}(\vec{n})
\right]^{\chi^{\uparrow}_{n_{t}}(\vec{n})}
\left[
b_{\downarrow}^{\dagger}(\vec{n})
\right]^{\chi^{\downarrow}_{n_{t}}(\vec{n})}
\right\}
 \, \ket{0}
  \label{eqn:transfer-matrix-0311}
\end{align}
where $\chi^{s}_{n_{t}}(\vec{n})$ counts the occupation number on each lattice site at time step $n_t$ and has values which are either 0 or 1. Let us define  the Grassmann functions,
\begin{align}
X(n_{t})
=_{}
\prod_{\vec{n}}
\left[
e^{\theta^{*}_{\uparrow}(n_{t},\vec{n})
   \theta_{\uparrow}(n_{t},\vec{n})}
   \,
e^{\theta^{*}_{\downarrow}(n_{t},\vec{n})
   \theta_{\downarrow}(n_{t},\vec{n})}
\right] \,,
 \label{eqn:transfer-matrix-0313}
\end{align}
and
\begin{equation}
M(n_t) =e^{-S_{H_{0}}[\theta,\theta^{*},n_{t}]}e^{-S_{V}[\theta,\theta^{*},n_{t}]}.
\end{equation}
From the relations given in Appendix~\ref{sec:append-operator-integral-formalism} the transfer matrix element between time steps $n_t$ and $n_t+1$ can be written in terms of these lattice Grassmann functions as
\begin{align}
\bra{\chi^{\uparrow}_{n_{t}+1}
,
\chi^{\downarrow}_{n_{t}+1}}
&
 \hat{M}
\ket{
\chi^{\uparrow}_{n_{t}}
,
\chi^{\downarrow}_{n_{t}}
}
\nonumber\\
&
=
\prod_{\vec{n}}
\left\{
\left[\frac{\overrightarrow{\partial}}{
\partial \theta_{\downarrow}^{*}(n_{t},\vec{n})}
\right]^{\chi^{\downarrow}_{n_{t}+1}(\vec{n})}
\left[\frac{\overrightarrow{\partial}}{
\partial \theta_{\uparrow}^{*}(n_{t},\vec{n})}
\right]^{\chi^{\uparrow}_{n_{t}+1}(\vec{n})}
\right\}
 X(n_{t}) \, M(n_{t})
 \nonumber\\
 &
 \times
\left.
\prod_{\vec{n}^{\prime}}
\left\{
\left[\frac{\overleftarrow{\partial}}{
\partial \theta_{\uparrow}(n_{t},\vec{n}^{\prime})}
\right]^{\chi^{\uparrow}_{n_{t}}(\vec{n}^{\prime})}
\left[\frac{\overleftarrow{\partial}}{
\partial \theta_{\downarrow}(n_{t},\vec{n}^{\prime})}
\right]^{\chi^{\downarrow}_{n_{t}}(\vec{n}^{\prime})}
\right\}\right|_{\tiny{\begin{array}{c}
                  \theta^{\prime}_{\uparrow}
                  =\theta^{\prime}_{\downarrow}=0             \\
                  \theta_{\uparrow}
                   =\theta_{\downarrow}=0 
                \end{array}}}
                \,.
\label{eqn:transfer-matrix-0316}
\end{align}
This result can be verified by checking the different possible combinations for the occupation numbers.  Because we have only one $\downarrow$ particle, the right-hand side is nonzero only if
\begin{align}
\sum_{\vec{n}} \chi^{\downarrow}_{n_{t}}(\vec{n})
=
\sum_{\vec{n}} \chi^{\downarrow}_{n_{t}+1}(\vec{n}) = 1\,.
\label{eqn:transfer-matrix-0317}
\end{align}

\begin{figure}[!htb]
\begin{center}
{\includegraphics[width=5in]{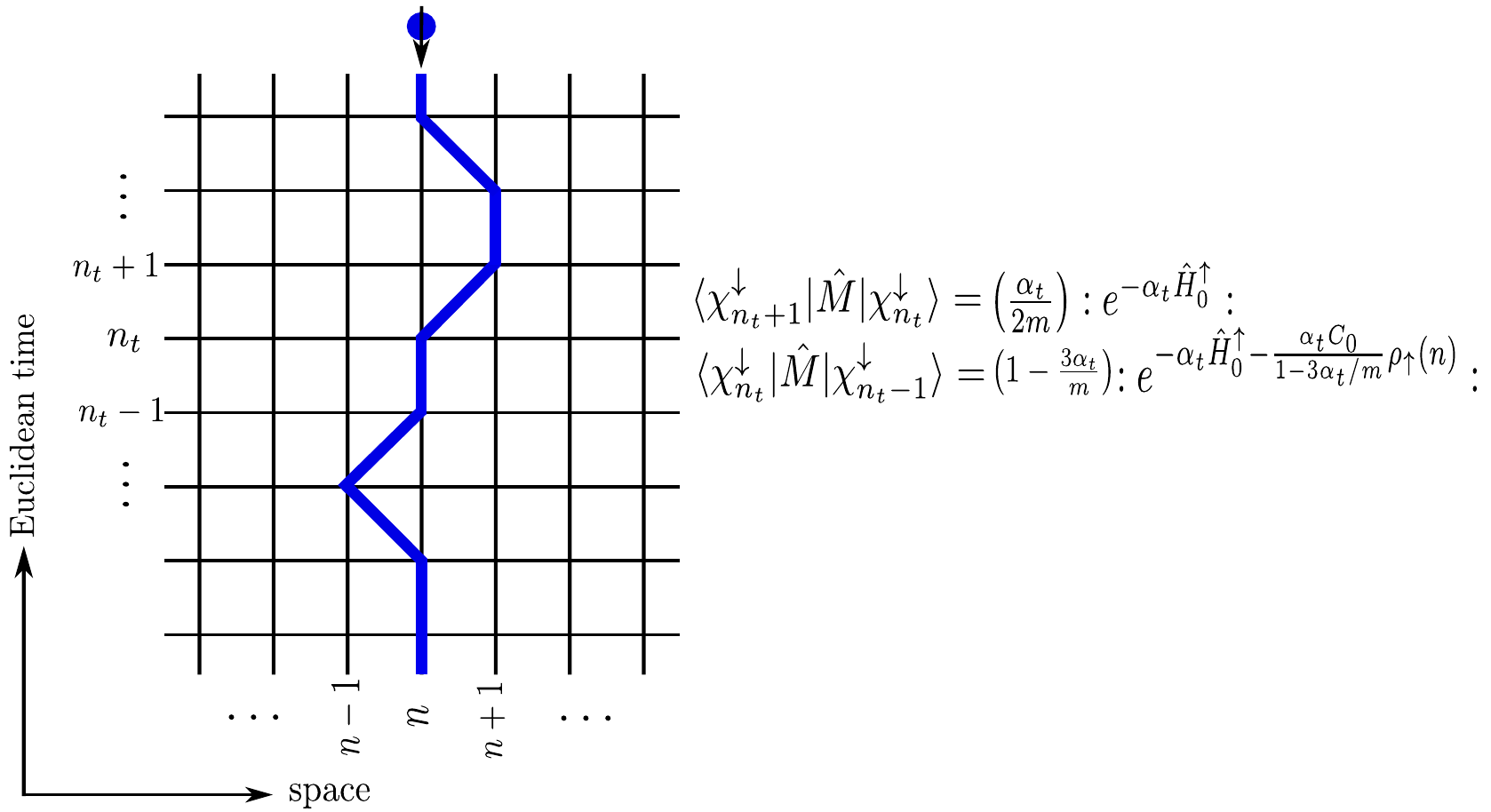}}
\end{center}
\caption{(Color online) A segment of a worldline configuration on a 1+1 dimensional Euclidean lattice.  See the main text for derivations of the reduced transfer-matrix operators.}%
\label{fig:worldline}
\end{figure}

We now derive the transfer matrix formalism for one spin-$\downarrow$ particle worldline in a medium consisting of an arbitrary number of spin-$\uparrow$ particles.  The impurity worldline is to be considered fixed. To provide  a simple visual representation of the worldine, we draw in Fig.~\ref{fig:worldline} an example of a single-particle worldline configuration on a 1+1 dimensional Euclidean lattice.

We now remove or ``integrate out" the impurity particle from the lattice action.  We consider first the case when the $\downarrow$ particle hops from $\vec{n}^{\prime \prime}$ to some nearest neighbor site.  In other words, $\chi^{\downarrow}_{n_{t}}(\vec{n}^{\prime \prime}) =1$ and
$\chi^{\downarrow}_{n_{t}+1}(\vec{n}^{\prime \prime}\pm\hat{l}) =1$ for some unit vector $\hat{l}$. In this case we have
\begin{align}
&\Braket{
\chi^{\uparrow}_{n_{t}+1},
\chi^{\downarrow}_{n_{t}+1}
|\hat{M}|
\chi^{\uparrow}_{n_{t}},
\chi^{\downarrow}_{n_{t}}
}
\nonumber\\
&
=
\prod_{\vec{n}}
\left\{
\left[\frac{\overrightarrow{\partial}}{
\partial \theta_{\uparrow}^{*}(n_{t},\vec{n})}
\right]^{\chi^{\uparrow}_{n_{t}+1}(\vec{n})}
\right\}
\slashed{X}(n_{t}) \,
\slashed{M}_{\vec{n}^{\prime\prime}\pm\hat{l},\vec{n}^{\prime\prime}}(n_{t})
\left.
\prod_{\vec{n}^{\prime}}
\left\{
\left[{\frac{\overleftarrow{\partial}}{
\partial \theta_{\uparrow}(n_{t},\vec{n}^{\prime})}}
\right]^{\chi^{\uparrow}_{n_{t}}(\vec{n}^{\prime})}
\right\}\right|_{\tiny{\begin{array}{c}
                  \theta^{\prime}_{\uparrow}
                  =0 \\
                  \theta_{\uparrow}
                  =0
                \end{array}}}
                \,,
\label{eqn:transfer-matrix-0319}
\end{align}
where
\begin{align}
\slashed{X}(n_{t})
=
\prod_{\vec{n}}
\left[
e^{\theta^{*}_{\uparrow}(n_{t},\vec{n})
   \theta_{\uparrow}(n_{t},\vec{n})}
\right] \,,
 \label{eqn:transfer-matrix-0321}
\end{align}
and\begin{align}
\slashed{M}_{\vec{n}^{\prime \prime}\pm\hat{l},\vec{n}^{\prime \prime}}(n_{t})
=
\left(
\frac{\alpha_{t}}{2 m}
\right)
 \exp
 \left\{
 -\alpha_{t} H^{\uparrow}_0[\theta_{s},\theta^*_{s},n_t]\right\}.
 \label{eqn:transfer-matrix-0323}
\end{align}
Next we consider the case when $\chi^{\downarrow}_{n_{t}}(\vec{n}^{\prime \prime})=1$ and $\chi^{\downarrow}_{n_{t}+1}(\vec{n}^{\prime \prime})=1$ which corresponds to no spatial hopping of the impurity worldline. Then we have
\begin{align}
&\Braket{
\chi^{\uparrow}_{n_{t}+1},
\chi^{\downarrow}_{n_{t}+1}
|\hat{M}|
\chi^{\uparrow}_{n_{t}},
\chi^{\downarrow}_{n_{t}}
}
\nonumber\\
&
=
\prod_{\vec{n}}
\left\{
\left[\frac{\overrightarrow{\partial}}{
\partial \theta_{\uparrow}^{*}(n_{t},\vec{n})}
\right]^{\chi^{\uparrow}_{n_{t}^{\prime}}(\vec{n})}
\right\}
\slashed{X}(n_{t}) \,
\slashed{M}_{\vec{n}^{\prime\prime},\vec{n}^{\prime\prime}}(n_{t})
\left.
\prod_{\vec{n}^{\prime}}
\left\{
\left[\frac{\overleftarrow{\partial}}{
\partial \theta_{\uparrow}(n_{t},\vec{n}^{\prime})}
\right]^{\chi^{\uparrow}_{n_{t}}(\vec{n}^{\prime})}
\right\}\right|_{\tiny{\begin{array}{c}
                  \theta^{\prime}_{\uparrow}
                  =0 \\
                  \theta_{\uparrow}
                  =0
                \end{array}}}
                \,,
 \label{eqn:transfer-matrix-029}
\end{align}
where\begin{align}
\slashed{M}_{\vec{n}^{\prime \prime},\vec{n}^{\prime \prime}}(n_{t})
=
\left(1- \frac{3\alpha_{t}}{m}\right)
&
 \exp
 \left\{
 -\alpha_{t} H^{\uparrow}_0[\theta_{s},\theta^*_{s},n_t]- \frac{\alpha_{t} C_{0}}{1- \frac{3\alpha_{t}}{m_{}}} \,\theta^*_{\uparrow}(n_{t},\vec{n}^{\prime\prime})\theta_{\uparrow}(n_{t},\vec{n}^{\prime\prime})\right\} \,.
 \label{eqn:transfer-matrix-033}
\end{align}

From these Grassmann lattice actions with the impurity integrated out, we can write down the corresponding transfer matrix operators.  When the impurity makes a spatial hop, the reduced transfer-matrix  operator is 
\begin{equation}
\hat{\slashed{M}}_{\vec{n}^{\prime \prime}\pm\hat{l},\vec{n}^{\prime \prime}}=
\left(
\frac{\alpha_{t}}{2 m}
\right ) \,  : \, \exp
 \left[
 -\alpha_{t} \hat{H}^{\uparrow}_{0}  \right]
 \, :.
 \label{eqn:t-matrix-stationary}
\end{equation}
When the impurity worldline remains stationary the reduced transfer-matrix operator is
\begin{equation}
\hat{\slashed{M}}_{\vec{n}^{\prime \prime},\vec{n}^{\prime \prime}}=
\left(1- \frac{3\alpha_{t}}{m}\right ) \,  : \, \exp
 \left[
 -\alpha_{t} \hat{H}^{\uparrow}_{0}-\frac{\alpha_{t} C_{0}}{1-\frac{3\alpha _{t}}{m}}
 \rho_{\uparrow}(\vec{n}^{\prime \prime})  \right]
 \, :.
  \label{eqn:t-matrix-hopping}
\end{equation}We note that these reduced transfer matrices are just one-body operators on the linear space of $\uparrow$ particles.

\section{Adiabatic Projection Method}
\label{sec:Adiabatic-Projection-Method}

In this section we describe our application of the adiabatic projection method using a set of cluster states constructed in momentum space. As already described in Eq.~(\ref{eqn:initialState-0001}) in a simplified notation, we let $\ket{\Psi_{\vec{p}}}$ be the fermion-dimer initial state  with relative momentum $\vec{p}$,
\begin{equation}
\lvert \Psi_{\vec{p}} \rangle = \tilde{b}^{\dagger}_{\uparrow}(-\vec{p})\hat{ M}^{L'_{t}}  \, \tilde{b}^{\dagger}_{\uparrow}(\vec{p})\tilde{b}^{\dagger}_{\downarrow}(\vec{0})\lvert 0 \rangle,
\label{eqn:initialState-0002}\
\end{equation} where we use the transfer matrix operator $\hat{ M}$ given in Eq.~(\ref{eqn:transfer-matrix-0200}) for some number of time steps $L'_{t}$.  The purpose of this time propagation is to allow the dimer to bind its constituents before injecting an additional $\uparrow$ particle.  In this part of the calculation we in fact increase the attractive interactions between the two spins to allow them to form the bound dimer faster.  We find that this trick increases the computational efficiency on large lattice systems. The dressed cluster states are defined as
\begin{align}
\ket{\Psi_{\vec{p}}}_{L_t/2} =  \hat{ M}^{L_t/2}\ket{\Psi_{\vec{p}}},
\end{align}
for some even number $L_t$, and the overlap between dressed cluster states is
\begin{align}
Z_{\vec{p}\vec{p}\,'}(L_t) 
= \prescript{}{L_t/2}{\braket{\Psi_{\vec{p}}|\Psi_{\vec{p}\,'}}}_{L_t/2} \,.
 \label{eqn:projection-amplitude-001}
\end{align}
We also calculate the matrix elements,
\begin{align}
Z_{\vec{p}\vec{p}\,'}(L_t+1) 
= \prescript{}{L_t/2}{\bra{\Psi_{\vec{p}}}}
\hat{M}
\ket{\Psi_{\vec{p}\,'}}_{L_t/2} \,.
\label{eqn:projection-amplitude-002}
\end{align}

For large $L_t$ we can obtain an accurate representation of the low-energy spectrum of $\hat{M}$ by defining the adiabatic transfer matrix as
\begin{align}
[\hat{M}^{a}(L_t)]_{\vec{p}\vec{p}\,'}
=
\sum_{\vec{p}\,''}Z^{-1}_{\vec{p}\vec{p}\,''}(L_t) \, \, Z_{\vec{p}\,''\vec{p}\,'}(L_t+1).\label{eqn:adibatic-hamitonian-0001} 
\end{align}
Alternatively we can also construct a symmetric version of the adiabatic transfer matrix as
\begin{align}
[\hat{M}^{a}(L_t)]_{\vec{p}\vec{p}\,'}
=
\sum_{\vec{p}\,'',\vec{p}\,'''}Z^{-1/2}_{\vec{p}\vec{p}\,''}(L_t) \, \, Z_{\vec{p}\,''\vec{p}\,'''}(L_t+1)Z^{-1/2}_{\vec{p}\,'''\vec{p}\,'}(L_t).\label{eqn:adibatic-hamitonian-0002}
\end{align}
Either form will produce exactly the same spectrum.  As with any transfer matrix, we interpret the eigenvalues $\lambda_i(L_t)$ of the adiabatic transfer matrix
as energies using the relations
\begin{equation}
e^{-E_i(L_t)\alpha_t} = \lambda_i(L_t), 
\quad  E_{i} (L_t)= -{\alpha_{t}}^{-1}\log \lambda_i(L_t).
\label{eqn:transient_energy} 
\end{equation}
The exact low-energy eigenvalues of the full transfer matrix $\hat{M}$ will be recovered in the limit $L_t \rightarrow \infty$.

As a special case, one can simply restrict the adiabatic projection calculation to  a single initial momentum state, for example, $\vec{p}=0$.  In that case the adiabatic transfer matrix is just the scalar ratio,
\begin{equation}
Z_{\vec{p}\vec{p}}(L_t+1)/Z_{\vec{p}\vec{p}}(L_t).
\end{equation}
However, we find that the energy calculations are significantly more accurate and converge much faster with increasing $L_t$ when using a set of several initial cluster states. 

\section{Impurity Monte Carlo Simulation}
\label{sec:simulation}

The reduced transfer matrices $\hat{\slashed{M}}_{\vec{n},\vec{n}'}$
in Eqs.~(\ref{eqn:t-matrix-stationary}) and (\ref{eqn:t-matrix-hopping}) are one-body operators on the linear space of $\uparrow$ particles.
Therefore, we can simply multiply the reduced transfer matrices together.  It is perhaps worthwhile to note that the ${\vec{n},\vec{n}'}$ subscripts are not the matrix indices of the reduced transfer matrix, but rather the coordinates of the $\downarrow$ particle that was integrated out.  The matrix indices of $\hat{\slashed{M}}_{\vec{n},\vec{n}'}$ are being left implicit.

The Euclidean time projection can be written  as a sum over worldline configurations of the $\downarrow$ particle. As a convenient shorthand we write \begin{equation}
\hat{\slashed{M}}_{\{\vec{n}_{j}\}}^{[L_t]}=
\hat{\slashed{M}}_{\vec{n}_{L_{t}},\vec{n}_{L_{t}-1}}
\ldots
\hat{\slashed{M}}_{\vec{n}_{1},\vec{n}_{0}},
\end{equation}
where $\vec{n}_{j}$ denotes the spatial position of the spin-$\downarrow$ particle at time step $j$.
 The projection amplitude for cluster states $\ket{\Psi_{\vec{p}}}$ and $\ket{\Psi_{\vec{p}\,'}}$ is, then,
 \begin{align}
Z_{\vec{p}\vec{p}\,'}(L_t) = \sum_{\vec{n}_{0},\ldots,\vec{n}_{L_{t}}}\bra{\Psi_{\vec{p}}}
\hat{\slashed{M}}_{\{\vec{n}_{j}\}}^{[L_t]}\ket{\Psi_{\vec{p}\,'}}.
 \label{eqn:Projected-matric-elements-010}
\end{align}
The states $\ket{\Psi_{\vec{p}\,}}$
and $\ket{\Psi_{\vec{p}\,'}}$ defined in Eq.~(\ref{eqn:initialState-0002}) are constructed using single-particle creation operators, and so the amplitude $Z_{\vec{p}\vec{p}\,'}(L_t)$ is just the determinant of a $2\times 2$ matrix of single-particle amplitudes.  As seen in Eq.~(\ref{eqn:initialState-0002}), there are  an extra $L'_t$ projection steps in between some of the creation operators.  This gives us the following structure, \begin{equation}
Z_{\vec{p}\vec{p}\,'}(L_t)=\sum_{\vec{n}_{0},\ldots,\vec{n}_{L_{t}}}\sum_{\vec{n}'_{0},\ldots,\vec{n}'_{L'_{t}}}\sum_{\vec{n}''_{0},\ldots,\vec{n}''_{L'_{t}}}\det M_{2\times2},
\end{equation} where $\vec{n}''_{0}=\vec{n}_{L_t}$,  $\vec{n}'_{L_t}=\vec{n}_{0}$, and\begin{equation}
M_{2\times2} = \begin{bmatrix}\langle  \vec p \rvert\hat{\slashed{M}}_{\{\vec{n}''_{j}\}}^{[L'_t]}\hat{\slashed{M}}_{\{\vec{n}_{j}\}}^{[L_t]}\hat{\slashed{M}}_{\{\vec{n}'_{j}\}}^{[L'_t]}\lvert \vec{p}\,' \rangle & \langle  \vec p \lvert\hat{\slashed{M}}_{\{\vec{n}''_{j}\}}^{[L'_t]}\hat{\slashed{M}}_{\{\vec{n}_{j}\}}^{[L_t]}\rvert- \vec{p}\,' \rangle \\
\langle-  \vec p \lvert\hat{\slashed{M}}_{\{\vec{n}_{j}\}}^{[L_t]}\hat{\slashed{M}}_{\{\vec{n}'_{j}\}}^{[L'_t]}\rvert \vec{p}\,' \rangle & \langle-  \vec p \lvert\hat{\slashed{M}}_{\{\vec{n}_{j}\}}^{[L_t]}\rvert- \vec{p}\,' \rangle \\
\end{bmatrix}.
\end{equation}The calculation of $Z_{\vec{p}\vec{p}\,'}(L_t)$ has now been recast as a problem of computing the determinant of the matrix $M_{2\times2}$ over all possible impurity worldlines.  We use a Markov chain Monte Carlo process to select worldline configurations.  The Metropolis algorithm is used to accept or reject configurations with importance sampling given by the weight function $\lvert Z_{\vec{p}\vec{p}}(L_t) \rvert$, where $\vec{p}$ is one of the initial momenta.   

We now benchmark our results for the low-energy spectrum calculated using adiabatic projection and the impurity Monte Carlo method. We compare these results with exact lattice results that we compute using the Lanczos iterative eigenvector method with a space of $\sim L^6$ basis states.  Although exact lattice results provide a useful benchmark test for the three-particle system, the extension to larger systems is computationally not viable because of exponential scaling in memory and CPU\ time.  In contrast, the impurity Monte Carlo calculation  does scale well to much larger systems.  In fact, many-body impurity systems are currently being studied in Ref.~\cite{Bour:2014}. 
\begin{table}[h]
\caption{Momentum of the dimer, $\vec{p}_{\rm{d}}$, with $p = 2\pi/L$. The total momentum of the  system is zero.}
\label{table:table-momenta0filling}
\centering
\begin{tabular}{c c}
  \hline\hline
  \, $n$ \, & $\vec{p}_{\rm{d}}$            \\ 
  \hline
  1      & $\Braket{p,0,0}$    \\
  2      & $\Braket{0,p,0}$     \\
  3      & $\Braket{0,0,p}$     \\
  4      & $\Braket{p,-p,0}$    \\
  5      & $\Braket{p,0,-p}$    \\
  6      & $\Braket{0,p,-p}$    \\
   \hline\hline
\end{tabular}
\end{table}

In our lattice calculations we take the particle mass to be the average nucleon mass, $938.92$ MeV, and the interaction strength $C_{0}$ is tuned to obtain the deuteron energy, $-2.2246$ MeV.   We use an $L^{3}$ periodic cubic volume with spatial lattice spacing $a = 1.97$ fm.  The values of $L$ used will be specified later.  In the temporal direction we use $L_t$ time steps with a temporal lattice spacing $a_{t} = 1.31$ fm/$c$.  

Let $N$ be the number of initial/final states.  We choose  the initial dimer momenta, $\vec{p}_{\rm{d}}$, as shown in Table~\ref{table:table-momenta0filling}.  In all cases the total momentum of the three-particle system is set to zero.  We label  and order the various possible dimer momenta with index $n=1,\cdots,N$.  We then construct the corresponding $N\times N$ adiabatic matrix, $[\hat{M}^{a}(L_t)]_{nn'}$, and obtain the $N$ low-lying energy states of the finite-volume system. There is no restriction on the choice of $N$.  Therefore, so long as the numerical stability of the matrix calculations is under control, it is advantageous to maximize the number $N$. While constructing a large adiabatic matrix requires more computational time, it significantly accelerates the convergence with the number of projection time steps, $L_t$. 
 
\captionsetup[subfigure]{position=bottom, labelfont=normalfont,textfont=normalfont,singlelinecheck=off,margin=30pt}
\begin{figure}[!ht]
        {\includegraphics[width=2.7in]{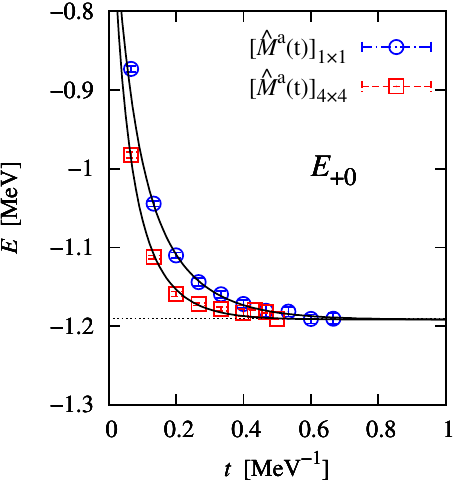}}
        \caption{(Color online) The ground-state energy is shown versus projection time $t$ using either one or four initial/final states. For comparison we show the exact lattice energies as dotted horizontal lines.}
        \label{fig:Energy1S}
\end{figure}

In Fig.~\ref{fig:Energy1S} we compare the ground-state energies  using $[\hat{M}^{a}(t)]_{1\times1}$ and $[\hat{M}^{a}(t)]_{4\times4}$ adiabatic matrices.
We are plotting the energies versus projection time $t=L_ta_t$.  The results shown are obtained using a lattice box of length $L\,a = 13.79$ fm, while the number of time steps is varied over a range of values to extrapolate to  the limit $t \to \infty$.  We use a simple exponential ansatz to extrapolate away the residual contribution from higher-energy states,\begin{equation}
E_i(t)=E_i(\infty)+c_ie^{-\Delta E_it}+\cdots. \label{eqn:error}
\end{equation}As can be seen clearly in the figure, the $[\hat{M}^{a}(t)]_{4\times4}$ results converge with a significantly faster exponential decay than the $[\hat{M}^{a}(t)]_{1\times1}$ results.  This is consistent with the derivation in Ref.~\cite{Pine:2013zja} that the energy gap $\Delta E_i$ in Eq.~(\ref{eqn:error}) is increased by including more initial states. The corresponding extrapolated ground-state energies obtained from the $[\hat{M}^{a}(t)]_{1\times1}$ and $[\hat{M}^{a}(t)]_{4\times4}$ adiabatic matrices are $-1.1918(46)$ MeV and  $-1.1916(25)$ MeV, respectively.

\captionsetup[subfigure]{position=bottom, labelfont=normalfont,textfont=normalfont,singlelinecheck=off,margin=30pt}
\begin{figure}[!ht]
        {\includegraphics[width=2.7in]{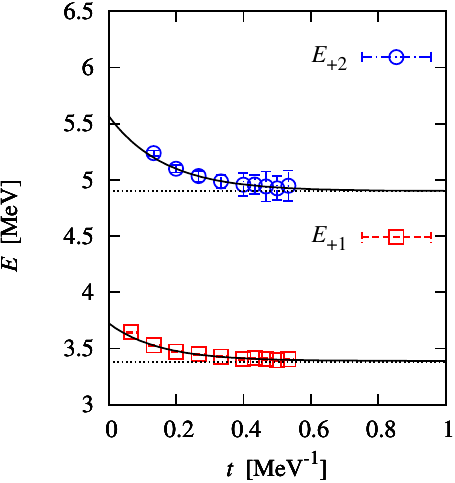}}
        \caption{(Color online) The first two excited state energies with even parity. For comparison we show the exact lattice energies as dotted horizontal lines.}
        \label{fig:Energy1D}
\end{figure}

In Figs.~\ref{fig:Energy1S} and \ref{fig:Energy1D} we plot the lowest lying even-parity energies as a function of Euclidean projection time $t$. To be able to calculate the two excited states in  Fig.~\ref{fig:Energy1D} we use the  $[\hat{M}^{a}(t)]_{4\times4}$ adiabatic matrix. For comparison the horizontal dotted lines in the plots represent the exact lattice energies obtained from the Lanczos iteration method.  The solid lines are exponential fits to the data using the ansatz in Eq.~(\ref{eqn:error}).  As seen from Figs.~\ref{fig:Energy1S} and \ref{fig:Energy1D} the corresponding extrapolated energies in Table~\ref{table:EvenParityEnergies}, we find that the calculations using adiabatic projection with impurity Monte Carlo are in excellent agreement with the exact lattice results. The energy level $E_{+1}$ has two-fold degeneracy. The degeneracies of this energy level are not shown in Fig.~\ref{fig:Energy1D} or Table~\ref{table:EvenParityEnergies}.

\begin{table}[h]
        \caption{The exact and Monte Carlo results for the ground-state and lowest lying even-parity energies in a periodic box of length $La=13.79$ fm. The Monte-Carlo results are obtained from the $[\hat{M}^{a}(t)]_{4\times4}$ adiabatic matrix.}
        \label{table:EvenParityEnergies}
        \centering
        \begin{tabular}{llll}
                \hline\hline
                \hspace{15mm}   & $E_{+0}$ (MeV) \hspace{10mm} & $E_{+1}$ (MeV)  \hspace{10mm} & $E_{+2}$ (MeV) \\ 
                \hline
                ${\rm{Exact}}$  & $-1.1904$ &  $3.3828$ & $4.9024$ \\
                $\rm{MC}$       &  $-1.1916(25)$& $3.3905(82)$  &  $4.9012(15)$\\
                \hline\hline
        \end{tabular}
\end{table}

\begin{figure}[!ht]
{\includegraphics[width=2.7in]{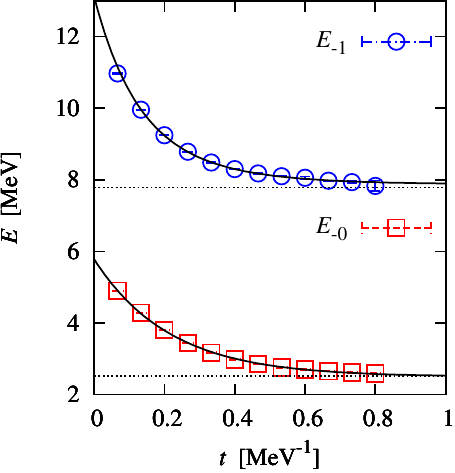}}
\caption{(Color online) The lowest two odd-parity energies as a function of Euclidean projection time $t$.  For comparison we show the exact lattice energies as dotted horizontal lines. }
\label{fig:EnergyOdd}
\end{figure}
In Fig.~\ref{fig:EnergyOdd} we present the energies for the lowest two states with odd-parity. To calculate these odd-parity energies we use five initial/final states and construct a $5\times5$ adiabatic matrix. The horizontal dotted lines  represent the exact lattice energies obtained from Lanczos iteration, and the solid lines are the exponential extrapolation fits.  We see from Fig.~\ref{fig:EnergyOdd} and Table~\ref{table:OddParityEnergies} again that the calculations using adiabatic projection with impurity Monte Carlo are in excellent agreement with the exact lattice results.   Both energy levels shown in Fig.~\ref{fig:EnergyOdd} have three-fold degeneracy. The degeneracies of these energy levels are not shown here.

\begin{table}[h]
\caption{The exact and Monte Carlo results for the energies of the lowest two odd-parity states in a periodic box of length $La=13.79$ fm. The Monte Carlo results are obtained from the $[\hat{M}^{a}(t)]_{5\times5}$ adiabatic matrix.}
\label{table:OddParityEnergies}
\centering
\begin{tabular}{l ll}
\hline\hline
        \hspace{15mm}        & $E_{-0}$ (MeV) \hspace{10mm} & $E_{-1}$ (MeV) \\\hline
${\rm{Exact}}$ & $2.509$ \qquad  \      &  $7.784$       \\\hline
$\rm{MC}$      &  $2.519(25)$  \qquad & $7.869(93)$ \\
\hline\hline
\end{tabular}
\end{table}

\section{Composite particles in finite volume}
In this section we present lattice results for the fermion-dimer elastic scattering phase shifts for angular momentum up to $\ell = 2$ using L\"uscher's finite-volume method.  As background for explaining the finite-volume calculations, we first briefly review L\"uscher's method.  Because the phase shifts depend crucially on an accurate calculation and analysis of finite-volume energy levels, we also discuss in the appendix some corrections which are from modifications of the dimer binding energy at finite volume. 

\subsection{L\"{u}scher's finite-volume method}

L\"{u}scher~\cite{Luscher1986105,Luscher1991531} has developed a well-known technique for extracting elastic phase shifts for two-body scattering from the volume dependence of two-body continuum states in a cubic periodic box. The method was extended to higher partial waves, two-body systems in moving frames, multichannel scattering cases, and scattering of particles with spin~\cite{Rummukainen1995397,PhysRevD.83.114508,PhysRevD.85.014506,PhysRevD.85.114507,PhysRevD.88.034502,PhysRevD.88.094507,PhysRevD.88.114507,PhysRevD.89.074507}. L\"{u}scher's framework was also successfully applied to the determination of resonance parameters~\cite{Bernard0806.4495}, and recently this technique was applied to moving frame calculations \cite{Eur.Phys.J.A48.114,PhysRevD.86.094513}. See Refs.~\cite{Bernard1010.6018,PhysRevD.85.014027,Doring1111.0616,PhysRevD.87.014502,Doring2013185} for further studies on the extraction of resonance properties at finite volume. We note also recent work on improving lattice interactions in effective field theories using L\"{u}scher's
method \cite{Endres:2011er}.

L\"{u}scher's relation between scattering phase shifts and two-body energy levels in a cubic periodic box has the following forms~\cite{Luscher1986105,Luscher1991531,PhysRevD.83.114508}:
\begin{equation}
p^{2\ell+1} \cot\delta_{\ell}(p)
=
\begin{dcases}
 \frac{2}{\sqrt{\pi}L}
 \mathcal{Z}_{0,0}(1;\eta) & \quad \text{for $\ell = 0$} \,,
\\ 
 \left(\frac{2\pi}{L}\right)^{3} \frac{\eta}{\pi^{3/2}}
 \mathcal{Z}_{0,0}(1;\eta) & \quad \text{for $\ell = 1$} \,,
\\ 
 \left(\frac{2\pi}{L}\right)^{5} \frac{1}{\pi^{3/2}}
 \left[
 \eta^{2}  \mathcal{Z}_{0,0}(1;\eta)
 +
 \frac{6}{7} \mathcal{Z}_{4,0}(1;\eta)
 \right] & \quad \text{for $\ell = 2$}  \,.
\end{dcases} 
\label{eqn:phaseshift-005}
\end{equation}
where
\begin{align}
\eta = \left(\frac{L p}{2\pi}\right)^{2}\,.
\end{align}
Here $\mathcal{Z}_{\ell,m}(1;\eta)$ are the generalized zeta functions~\cite{Luscher1986105,Luscher1991531},
\begin{align}
\mathcal{Z}_{\ell,m}(1;\eta)
=
\sum_{\vec{n}}
\frac
{|\vec{n}|^{\ell} \, \text{Y}_{\ell,m}(\hat{n})}
{|\vec{n}|^{2} - \eta}
\,,
\label{eqn:phaseshift-009}
\end{align}
and $Y_{\ell,m}(\hat{n})$ are the spherical harmonics. We can evaluate the zeta functions using exponentially accelerated expressions \cite{PhysRevD.83.114508}.  For $\ell,m= 0$ we have
\begin{align}
\mathcal{Z}_{0,0}(1;\eta)
=
\pi e^{\eta}(2\eta -1) 
&+ \frac{e^{\eta}}{2\sqrt{\pi}}\sum_{\vec{n}}
\frac{e^{-|\vec{n}|^{2}}}{|\vec{n}|^{2}-\eta}
\nonumber\\
&-\frac{\pi}{2}\int_{0}^{1} d\lambda\frac{e^{\lambda \eta}}{\lambda^{3/2}}
\left(4\lambda^{2}\eta^{2} -
 \sum_{\vec{n}}e^{-\pi^{2}|\vec{n}|^{2}/\lambda}\right)
\,,
\label{eqn:phaseshift-031}
\end{align}
 and for arbitrary $\ell$ and $m$,
 \begin{align}
 \mathcal{Z}_{\ell,m}(1;\eta)
 =
\sum_{\vec{n}}
&
\frac
{|\vec{n}|^{\ell} \, \text{Y}_{\ell,m}(\hat{n})}
{|\vec{n}|^{2} - \eta}
e^{-\Lambda(|\vec{n}|^{2} - \eta)} 
\nonumber\\
 &+\int_{0}^{\Lambda} d\lambda
 \left(\frac{\pi}{\lambda}\right)^{\ell+3/2}
 e^{\lambda \eta}
\sum_{\vec{n}}
\frac
{|\vec{n}|^{\ell} \, \text{Y}_{\ell,m}(\hat{n})}
{|\vec{n}|^{2} - \eta}
e^{-\pi^{2}|\vec{n}|^{2}/\lambda}
 \,.
 \label{eqn:phaseshift-035}
 \end{align}

\subsection{Results for the elastic phase shifts}

\captionsetup[subfigure]{position=bottom, labelfont=normalfont,textfont=normalfont,singlelinecheck=off,margin=30pt}
\begin{figure}[h]
{\includegraphics[width=3.2in]{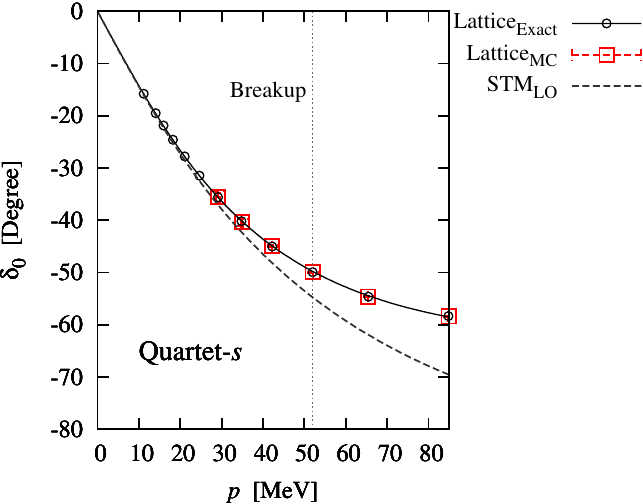}}
\caption{(Color online) The \textit{s}-wave scattering phase shift versus the relative momentum between fermion and dimer.}
\label{fig:s-wavePhaseshift}
\end{figure}
\captionsetup[subfigure]{position=bottom, labelfont=normalfont,textfont=normalfont,singlelinecheck=off,margin=30pt}
\begin{figure}[h]
{\includegraphics[width=3.2in]{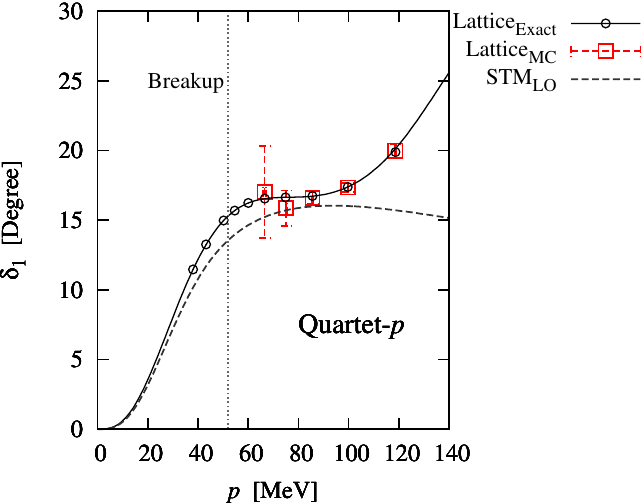}}
\caption{(Color online) The \textit{p}-wave scattering phase shift versus the relative momentum between fermion and dimer.}
\label{fig:p-wavePhaseshift}
\end{figure}
\captionsetup[subfigure]{position=bottom, labelfont=normalfont,textfont=normalfont,singlelinecheck=off,margin=30pt}
\begin{figure}[h]
{\includegraphics[width=3.2in]{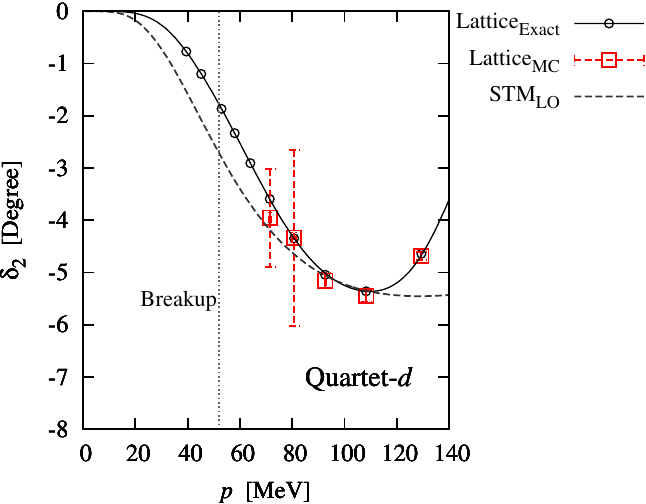}}
\caption{(Color online) The \textit{d}-wave scattering phase shift versus the relative momentum between fermion and dimer.}
\label{fig:d-wavePhaseshift}
\end{figure}

We now use our lattice results for the finite-volume energies and use Eq.~(\ref{eqn:phaseshift-005})  to determine the elastic phase shifts.  We compute phase shifts using data from the impurity Monte Carlo calculations as well as the exact lattice energies using the Lanczos method.  The fermion-dimer system that we are considering corresponds exactly to neutron-deuteron scattering in the spin-quartet channel at the leading order in pionless effective field theory. Therefore we choose to set the dimer energy to the physical deuteron energy of $-2.2246$ MeV and compare our lattice results to calculations of neutron-deuteron scattering in the continuum and infinite-volume limits at leading order in pionless effective field theory \cite{Bedaque2000357,Gabbiani2000601,Rupak200373}.  The Skorniakov-Ter-Martirosian (STM) integral equation for the $T$-matrix is
\begin{align}
T_{\ell}(k,p) = - & \frac{8 \pi \gamma}{m p k}
\,
Q_{\ell}\left(\frac{p^{2}+k^{2}-mE-i0^{+}}{pk}\right)
\nonumber\\
&
-\frac{2}{\pi}\int_{0}^{\infty} dq \, \frac{q}{p} \frac{T_{\ell}(k,q)}{\sqrt{3q^{2}/4-mE-i0^{+}}-\gamma}
\,
Q_{\ell}\left(\frac{p^{2}+q^{2}-mE-i0^{+}}{pq}\right),
\end{align}
where $\gamma$ is the dimer binding energy, $E = 3p^{2}/(4m)-\gamma^{2}/m$ is the total energy, and $Q_{\ell}$ is the Legendre function of the second kind,
\begin{align}
Q_{\ell}(a) = \frac{1}{2}\int_{-1}^{1}
dx \, \frac{P_{\ell}(x)}{x+a}\,.
\end{align}
The scattering phase shifts can be calculated from the on-shell $T$-matrix formula,
\begin{align}
T_{\ell}(p,p) = \frac{3\pi}{m} \frac{p^{2\ell}}{p^{2\ell+1} \cot\delta_{\ell} - ip^{2\ell+1}}\,.
\end{align}

We show results for the $s$-wave, $p$-wave, and $d$-wave phase shifts in Figs.~\ref{fig:s-wavePhaseshift}, \ref{fig:p-wavePhaseshift}, and \ref{fig:d-wavePhaseshift}, respectively.  The square points  indicate the data from the lattice Monte Carlo simulations, the circular points are the exact lattice calculations, and the solid lines are a fit of the exact lattice data using an effective range expansion, 
\begin{align}
  p^{2\ell+1} \cot\delta_{\ell}(p)
  = -\frac{1}{a_{\ell}}+\frac{1}{2}r_{\ell} \,  p^{2}+\mathcal{O}(p^{4}) \,.
 \label{eqn:append:phaseshift-001}
\end{align}
The dashed lines are leading order results from the STM calculation.  The dotted vertical lines indicate the inelastic breakup threshold of the dimer. The range of lattice box sizes is $L \leq 16$ for the exact lattice and $L \leq 9$ for the Monte Carlo calculations. As it is demonstrated in Figs.~\ref{fig:s-wavePhaseshift} - \ref{fig:d-wavePhaseshift}, this range of the box sizes allows us to access the energy regime down to approximately 10 MeV for the $s$ wave and $40$ MeV for the $p$ wave and $d$ wave.

Where we have overlapping data, we find excellent agreement between the Monte Carlo and exact lattice phase shifts.  At very low energies we find that Monte Carlo calculations of the phase shifts become impractical because the high sensitivity of L{\"u}scher's method upon small deviations in the finite-volume energies.  This should be regarded more as a limitation of L{\"u}scher's formalism rather than a deficiency of the adiabatic projection method or impurity Monte Carlo.

We also find quite good agreement between the STM continuum results and the lattice results for the $s$-wave and $p$-wave phase shifts.  Below the inelastic breakup threshold, the small deviation can be regarded as an estimate of lattice discretization errors. For the $s$-wave and $p$-wave phase shifts the eigenstates of the angular momentum in the $\mathrm{SO}(3)$ rotational symmetry of continuum space decompose into the irreducible representations $A_1$ and $T_1$ of the cubic rotational group, respectively. Therefore, there is one-to-one correspondence between the finite-volume energy spectrum and the phase shifts. However, the eigenstates of the angular momentum for $\ell =2$ decompose into the irreducible representations $E$ and $T_2$ of the cubic rotational group. In our analysis for the $d$ wave  we consider only the irreducible representation $E$ of the cubic rotational group, and we believe this is the origin of the larger deviation in the $d$-wave phase shifts.

Above the breakup threshold there are also systematic errors because our analysis using L\"uscher's method does not account for inelastic breakup processes.  Nevertheless we see that the agreement with the STM calculation  for the elastic phase shifts is still quite good, and the STM\ calculation does correctly account for breakup effects.  To our knowledge these results are the first lattice calculations of fermion-dimer scattering in the $p$- and $d$-wave channels. 

\begin{table}[h]
        \caption{The exact and Monte Carlo results for the energies up to $L=9$ used in the $s$-wave, $p$-wave, and $d$-wave phase shift calculations given in Figs.~\ref{fig:s-wavePhaseshift}, \ref{fig:p-wavePhaseshift}, and \ref{fig:d-wavePhaseshift}.}
        \label{table:AllEnergies}
        \centering
        \begin{tabular}{r r r r r r r}
                \hline\hline
                \multicolumn{1}{c}{$L$}
                &  \multicolumn{2}{c}{$E_{+0}$ (MeV)}
                &  \multicolumn{2}{c}{$E_{-0}$ (MeV)} 
                &  \multicolumn{2}{c}{$E_{+1}$ (MeV)}
                \\ \cmidrule[0.4pt](lr{0.5em}){2-3}
                \cmidrule[0.4pt](lr{0.5em}){4-5}%
                \cmidrule[0.4pt](lr{0.5em}){6-7}%
                &\multicolumn{1}{c}{${\rm{Exact}}$}
                &\multicolumn{1}{c}{${\rm{MC}}$}       
                & \multicolumn{1}{c}{${\rm{Exact}}$}  
                & \multicolumn{1}{c}{${\rm{MC}}$}       
                & \multicolumn{1}{c}{${\rm{Exact}}$}
                & \multicolumn{1}{c}{${\rm{MC}}$}
                \\\hline
                ~4~
                & $1.646$
                & $1.649(7)$      
                & $10.208$          
                & $10.259(154)$      
                & $13.127$          
                & $13.139(13)$ 
                \\
                ~5~  
                & $0.164$          
                & $0.161(3)$      
                & $6.401$          
                & $6.395(30)$      
                & $8.254$          
                & $8.259(10)$ 
                \\
                ~6~  
                & $-0.684$          
                & $-0.634(3)$     
                & $4.037$          
                & $4.038(13)$      
                & $5.285$          
                & $5.293(10)$ 
                \\
                ~7~  
                & $-1.190$          
                & $-1.192(3)$     
                & $2.509$          
                & $2.512(15)$      
                & $3.383$          
                & $3.390(10)$ 
                \\
                ~8~  
                & $-1.506$          
                & $-1.503(11)$      
                & $1.475$          
                & $1.493(30)$      
                & $2.104$          
                & $2.105(90)$ 
                \\
                ~9~  
                & $-1.711$          
                & $-1.706(13)$      
                & $0.746$          
                & $0.737(62)$      
                & $1.209$         
                & $1.224(40)$ 
                \\\hline\hline
        \end{tabular}
\end{table}

In Table~\ref{table:AllEnergies} we present the energies obtained from the exact and Monte Carlo calculations up to $L=9$. These fermion-dimer energies are used in the $s$-wave, $p$-wave and $d$-wave phase shift calculations. As can be seen from the table the relative errors in the energies are relatively larger for higher partial waves. Therefore, this explains, together with the fact that the L\"uscher method becomes more difficult at higher angular momentum, the larger error bars in the $p$-wave and $d$-wave phase shifts given in Figs.~\ref{fig:p-wavePhaseshift} and \ref{fig:d-wavePhaseshift}.

\section{Summary and discussion}

In this paper we have presented the adiabatic projection method and its first application using Monte Carlo methods.  The adiabatic method is a general framework for studying scattering and reactions on the lattice. The method constructs a low-energy effective theory for clusters, and in the limit of large Euclidean projection time the description becomes exact. In previous studies~\cite{Rupak:2013aue,Pine:2013zja} the initial cluster states were parameterized by the initial spatial separations between clusters. In this study we have used a new technique which parameterizes the cluster states according to the relative momentum between clusters. This new approach is crucial for doing calculations with a small number of initial states to improve the efficiency of the Monte Carlo calculations.
 The system we have analyzed in detail here is fermion-dimer elastic scattering for two-component fermions interacting via zero-range attractive interactions.

For our calculations we have introduced a new Monte Carlo algorithm which we call impurity lattice Monte Carlo. This can be seen as a hybrid algorithm in between worldline and auxiliary-field Monte Carlo simulations. In impurity Monte Carlo we use worldline Monte Carlo for the impurities, and these impurity worldlines are acting as additional auxiliary fields in the simulation of the other particles.  By using the impurity lattice Monte Carlo algorithm, we have found significant improvement over more standard auxiliary-field Monte Carlo  calculations.  In addition to greater speed and efficiency of the calculations, we also found a reduction of fermonic sign oscillations, and this has greatly improved the resulting accuracy.

We have found that the adiabatic projection method with impurity Monte Carlo enables highly accurate calculations of the finite-volume energy levels of the fermion-dimer system. From these energy levels we have used L\"uscher's method to present the first lattice calculations of $p$-wave and $d$-wave phase shifts for fermion-dimer elastic scattering. In addition to finding excellent agreement between Monte Carlo and exact lattice phase shifts, we have also found good agreement with continuum STM calculations of neutron-deuteron elastic scattering in the spin-quartet channel at leading order in pionless effective field theory. 

Our results show that the adiabatic projection method with Monte Carlo simulations is a viable approach to calculating elastic phase shifts.  The method can be applied in a straightforward manner to other two-cluster scattering systems.  One area where more work is needed is that our application of L\"uscher's method does not account for inelastic breakup processes.  Another area that needs improvement is that L\"uscher's method has too much sensitivity to small changes
in the finite-volume energy levels.  For these reasons we are now working to develop new methods which incorporates more information from the adiabatic projection wave function to extract scattering information in a more robust manner.

\acknowledgments
We thank Gautam Rupak for extensive discussions on several aspects of the project and for help on the large-scale computational runs.  We are also thank Shahin Bour, Hans-Werner Hammer and Ulf-G. Mei{\ss}ner for discussions on the impurity Monte Carlo method.  We are also grateful to Harald Grie{\ss}hammer for helpful correspondence regarding the STM\ calculations. The computer supports provided by the NCSU High
Performance Computing center.
Financial support provided by U.S. Department of Energy Grant No.DE-FG02-03ER41260.

\appendix

\section{The exact relations between the normal-ordered lattice transfer matrix and lattice Grassmann integrals}
\label{sec:append-operator-integral-formalism}
In this appendix we give some basic relations between the normal-ordered lattice transfer matrix and lattice Grassmann integrals. Materials presented here are useful to carry out the derivations in Sections~\ref{sec:LatticePIandTM} and \ref{sec:ImpurityLMC}.

We start with the annihilation and creation operator, $b$ and $b^\dagger$, for one fermion species on a single lattice site.  We use the usual occupation number basis, $\ket{0}$ and $\ket{1}$, where\begin{equation}
b\ket{0} = 0, \; \ket{1}=b^\dagger\ket{0}, \; \braket{0|0}=\braket{1|1}=1.
\end{equation}Let us also consider two anticommuting Grassmann variables $\theta$ and $\theta^*.$   The matrix elements of any normal-ordered operator $:f(b,b^\dagger):$ can be calculated in terms of the Grassmann variables as
\begin{equation}
 \bra{i}:f(b,b^\dagger):\ket{j}= \left\{ \left[ \frac{\overrightarrow{\partial}}{\partial \theta^*} \right]^ie^{\theta^*\theta}f(\theta,\theta^*)\left[ \frac{\overleftarrow{\partial}}{\partial
\theta} \right]^j \right\}_{\theta=\theta^*=0}\label{eqn:derivatives}
\end{equation}
for indices $i,j = 0$ or $1$.  This identity is easily checked by considering
all possible normal-ordered functions of $b$ and $b^\dagger$. It is straightforward to check that this identity can be generalized to an arbitrary number of fermion operators.  This then leads us to Eq.~(\ref{eqn:transfer-matrix-0316}).

We now define the Grassmann integrals,  \begin{equation}
\int d\theta=\int d\theta^*=0, \; \int d\theta \theta =\int d\theta^* \theta^*=1.
\end{equation}  The equivalence between Grassmann integration and Grassmann differentiation gives 

\begin{equation}
\bra{i}:f(b,b^\dagger):\ket{j}=\int \left( d\theta^* \right)^i e^{\theta^*\theta}f(\theta,\theta^*)\left(- d\theta \right)^j. 
\end{equation}
 \\ Let us now consider the matrix elements of a product of normal-ordered operators,
\begin{equation}
\bra{i}:f_1(b,b^\dagger):\,:f_{0}(b,b^\dagger):\ket{j}=\sum_{k}\bra{i}:f_1(b,b^\dagger):\ket{k}\bra{k}:f_{0}(b,b^\dagger):\ket{j}.
\end{equation}
From Eq.~(\ref{eqn:derivatives}) we can write the right-hand side as\begin{equation}
\sum_k\left\{ \left[ \frac{\overrightarrow{\partial}}{\partial
\theta_1^*} \right]^ie^{\theta_1^*\theta_1}f(\theta_1,\theta_1^*)\left[ \frac{\overleftarrow{\partial}}{\partial
\theta_1} \right]^k  \left[ \frac{\overrightarrow{\partial}}{\partial
\theta_0^*} \right]^k e^{\theta_0^*\theta_0}f(\theta_0,\theta_0^*)\left[ \frac{\overleftarrow{\partial}}{\partial
\theta_0} \right]^j \right\}_{\theta_{0,1}=\theta^*_{0,1}=0}.
\end{equation}
We note that for any functions $g_0$ and $g_1$,
\begin{equation}
\sum_k\left\{ g_{1}(\theta_1)\left[ \frac{\overleftarrow{\partial}}{\partial
\theta_1} \right]^k  \left[ \frac{\overrightarrow{\partial}}{\partial
\theta_0^*} \right]^k g_{0}(\theta_0^*) \right\}_{\theta_{1}=\theta^*_{0}=0}=-\int d\theta_1d\theta^*_0e^{-\theta^*_0\theta_1}g_1(\theta_1)g_0(\theta^*_0).
\end{equation}
Therefore we have\begin{equation}
\bra{i}:f_1(b,b^\dagger):\,:f_{0}(b,b^\dagger):\ket{j}=-\int (d\theta_1^*)^id\theta_1d\theta^*_0e^{\theta_{1}^*\theta_{1}}f_{1}(\theta_1,\theta_1^*)e^{\theta_0^*(\theta_0-\theta_1)}f_{0}(\theta_0,\theta_0^*)(-d\theta_0)^j.
\end{equation}
More generally, for the product of $L_t$ normal-ordered operators, we find
\begin{align}
\bra{i}: f_{n-1}(b,b^\dagger):\cdots:f_{0}(b,b^\dagger):\ket{j} &= (-1)^{n-1} \int (d\theta_{n-1}^*)^ie^{\theta_{n-1}^*\theta_{n-1}}f_{n-1}(\theta_{n-1},\theta^*_{n-1}) \nonumber \\ & \times \prod_{m=0}^{n-2} [d\theta_{m+1}d\theta^*_m e^{\theta_{m}^*(\theta_{m}-\theta_{m+1})}f_{m}(\theta_{m},\theta^*_m)](-d\theta_0)^j,  \label{eqn:matrixelement} 
\end{align}
where the product on the right-hand side is a time-ordered product, although the ordering will not matter in physical applications where the functions $f_{m}$ are even in the Grassmann variables.

We note that for any function $g(\theta_{0},\theta^*_{n-1})$, 

\begin{equation}
\sum_i \int (d\theta_{n-1}^*)^i g(\theta_{0},\theta^*_{n-1})(-d\theta_0)^i =\int d\theta_{0} d\theta_{n-1}^*e^{\theta^*_{n-1}\theta_0}g(\theta_{0},\theta^*_{n-1}).  
\end{equation}
From this and Eq.~(\ref{eqn:matrixelement}), we derive the trace identity,
\begin{align}
\Tr \left[: f_{n-1}(b,b^\dagger):\cdots:f_{0}(b,b^\dagger): \right] &= 
\int  \prod_{m=0}^{n-1} [d\theta_{m}d\theta^*_m e^{\theta_{m}^*(\theta_{m}-\theta_{m+1})}f_{m}(\theta_{m},\theta^*_m)],
\end{align}
where $\theta_{n}=-\theta_0$ and the factors of $-1$ vanish upon reordering the integration measure.  The straightforward generalization of these results to more than one fermion annihilation and creation operator leads to Eq.~(\ref{eqn:G-path-T-matrix-001}).  

\section{Finite-volume binding energy corrections and topological volume corrections for scattering with arbitrary $\ell$}

\label{sec:append:topological-phases}
To apply L\"uscher's finite-volume method with maximal accuracy, we consider also finite-volume
corrections to the binding energy of the dimer.  The finite-volume correction to two-body $s$-wave binding energies was derived in Refs.~\cite{Luscher:1985dn} and extended to arbitrary angular momentum in Ref.~\cite{Konig:2011nz,Konig:2011ti}.  There was also significant work towards understanding three-body binding energy corrections at finite volume \cite{Kreuzer:2010ti,Kreuzer:2012sr}.   

It was noticed in Ref.~\cite{PhysRevD.84.091503} that the finite-volume corrections to the dimer binding energy is dependent on the motion of the dimer.  This fact was used to cancel out finite-volume corrections to the binding energy \cite{Davoudi:2011md}.  The dimer motion induces phase-twisted boundary conditions on the dimer's relative-coordinate wave function.  These effects are called topological volume corrections and were found to have an effect on the finite-volume analysis for scattering of the dimer.  The study of topological volume corrections were carried out for $s$-wave scattering in Refs.~\cite{PhysRevD.84.091503,PhysRevC.86.034003} and further applied in Refs.~\cite{A.Rokash2013.3386,Pine:2013zja}.  In the following we show the extension to general partial wave $\ell$. 

The general solution of the Helmholtz equation has the form of
\begin{align}
\psi_{p}(\vec{r}) = 
  \sum_{\ell,m} c_{\ell,m}(p) \, G_{\ell,m}(\vec{r},p^{2}) \,.
  \label{eqn:topologicalphases-000}
\end{align}
The functions $G_{\ell,m}(\vec{r},p^{2})$ form a linearly independent complete basis set and are defined as
\begin{align}
G_{\ell,m}(\vec{r},p^{2}) = \mathcal{Y}_{\ell,m}(\nabla)
\, G(\vec{r},p^{2})
\,.
\label{eqn:topologicalphases-001}
\end{align}
Here $\mathcal{Y}_{\ell,m}$ are the solid spherical harmonic polynomials and defined in terms of the spherical harmonics as
\begin{align}
\mathcal{Y}_{\ell,m}(\vec{r})
=
r^{\ell} \, \text{Y}_{\ell,m}(\theta,\phi)
\,,
\label{eqn:topologicalphases-005}
\end{align}
and $G(\vec{r},p^{2})$ is the periodic Green's function solution to the Helmholtz equation for $\ell,m= 0$,
\begin{align}
G_{0,0}(\vec{r},p^{2})
=
G(\vec{r},p^{2})
=
\frac{1}{L^{3}}
\sum_{\vec{k}}
\frac
{e^{\frac{2i\pi}{L}\vec{k}\cdot\vec{r}}}
{\left(\frac{2\pi}{L}\vec{k}\right)^{2} - p^{2}}
\,.
\label{eqn:topologicalphases-011}
\end{align}
Using Eq.~(B1) of Ref.~\cite{Luscher1991531}, we have
\begin{align}
G_{\ell,m}(\vec{r},p^{2}) = r^{\ell} \, \text{Y}_{\ell,m}(\theta,\phi)
\left(\frac{1}{r}\frac{\partial}{\partial r}\right)^{\ell}
\, G(\vec{r},p^{2})
\,.
\label{eqn:topologicalphases-009}
\end{align}
Inserting Eq.~(\ref{eqn:topologicalphases-011}) into Eq.~(\ref{eqn:topologicalphases-009}), we write the asymptotic form of the scattering wave function as
\begin{align}
u_{\ell}(r) = 
C
\,
\sum_{\vec{k}}
|\vec{k}|^{\ell}
\frac
{e^{\frac{2i\pi}{L}\vec{k}\cdot\vec{r}}}
{\left(\frac{2\pi}{L}\vec{k}\right)^{2} - p^{2}}
\,,
\label{eqn:topologicalphases-015}
\end{align}
where $C$ is the normalization coefficient. The derivation of the topological volume corrections for the $s$-wave scattering of two composite particles $A$ and $B$ is given in Ref.~\cite{PhysRevD.84.091503,PhysRevC.86.034003,A.Rokash2013.3386}. Here we focus on the fermion-dimer scattering and derive the topological volume corrections for higher partial waves. 

In this analysis we take the continuum limit.  We let the total momentum of the fermion plus dimer system to be zero and let $p$ be the magnitude of the relative momentum between the fermion and dimer.  Let $E_{\text{d},\vec{0}}(\infty)$ be the dimer energy at infinite volume and $m_{\rm{d}}$ be the dimer mass. Then the fermion-dimer energy at infinite volume, $E_{\text{df}}(p,\infty)$, is
\begin{align}
E_{\text{df}}(p,\infty) = \frac{p^{2}}{2 m_{\text{d}}} + \frac{p^2}{2m} + E_{\text{d},\vec{0}}(\infty) \,.
\label{eqn:fermion-dimer-energy-021}
\end{align}
As in previous studies of fermion-dimer scattering on the lattice \cite{PhysRevD.84.091503,PhysRevC.86.034003,A.Rokash2013.3386,Pine:2013zja}, we calculate the effective dimer mass of the dimer on the lattice by computing the dimer dispersion relation. Now we let $E_{\text{df}}(p,L)$ be the finite-volume energy of the fermion-dimer system.  Following  Refs.~\cite{PhysRevD.84.091503,PhysRevC.86.034003,A.Rokash2013.3386}, we can compute the expectation value,
\begin{align}
E_{\text{df}}(p,L) = \frac{\int d^3r \, u_{\ell}^{*}(r) \hat{H}  u_{\ell}(r)}{\int d^3r \,|u_{\ell}(r)|^{2}}
= 
\frac{1}{\mathcal{N}_{\ell}} \sum_{\vec{k}}^{{k}_{max}}
|\vec{k}|^{2\ell}
 \frac{\frac{p^{2}}{2m_{\text{d}}}+ \frac{p^2}{2m}
  +E_{\text{d},\vec{k}}(L)}{\left(\vec{k}^{2} -\eta\right)^{2}}
  \,,
\label{eqn:topologicalphases-019}
\end{align}
where $E_{\text{d},\vec{k}}(L)$ is the finite-volume energy of the dimer with momentum $\vec{k}$,
$\mathcal{N}_{\ell}$ is defined as \begin{align}
\mathcal{N}_{\ell} = \sum_{\vec{k}}^{{k}_{max}}
|\vec{k}|^{2\ell}
\left(\vec{k}^{2} -\eta\right)^{-2},
\end{align} and  $\eta = \left(\frac{Lp}{2 \pi}\right)^{2}$.  For $\ell > 0$ the summations are divergent and we must cut off the short distance behavior at some momentum scale $\Lambda$ characterizing the range of the fermion-dimer interactions.  The corresponding maximum index value $k_{max}$ scales as $\Lambda L/(2 \pi)$. 

Let $\Delta E_{\text{d},\vec{0}}(L) = E_{\text{d},\vec{0}}(L) - E_{\text{d},\vec{0}}(\infty)$ be the finite-volume energy shift of the dimer energy in its rest frame, and $\Delta E_{\text{d},\vec{k}}(L) = E_{\text{d},\vec{k}}(L)- E_{\text{d},\vec{k}}(\infty)$ be the finite-volume energy shift of the dimer energy with momentum $\vec{k}$.  One can show that \cite{PhysRevD.84.091503},
\begin{align}
\frac{ \Delta E_{\text{d},\vec{k}}(L)}{\Delta E_{\text{d},\vec{0}}(L)} 
 =
 \frac{1}{3} \sum_{i =1}^{3} \cos\left(2\pi k_{i} \, \alpha \right)\, .
\label{eqn:dimer_vol_corr}
\end{align}
Using Eqs.~(\ref{eqn:fermion-dimer-energy-021}), (\ref{eqn:topologicalphases-019}), and (\ref{eqn:dimer_vol_corr}), we can now write the fermion-dimer energy correction at finite volume as
\begin{align}
E_{\text{df}}(p,L) -E_{\text{df}}(p,\infty) 
=
\tau_{\ell}(\eta) \, \Delta E_{\text{d},\vec{0}}(L) \,,
\label{eqn:topologicalphases-025}
\end{align}
where $\tau_{\ell}(\eta)$ is the topological factor,
\begin{align}
\tau_{\ell}(\eta)
= 
\frac{1}{\mathcal{N}_{\ell}} \sum_{\vec{k}}^{{k}_{max}}
 \frac{|\vec{k}|^{2\ell} \, 
 \sum_{i =1}^{3} \cos\left(2\pi k_{i} \, \alpha \right)}{3\left(\vec{k}^{2} -\eta\right)^{2}}
\,,
\label{eqn:topologicalphases-029}
\end{align}
with $\alpha = m_{}/(m+m_{\rm{d}})=1/3$. 
Because of the short distance  behavior of the momentum mode summations for $\ell > 0$, we find that the topological phase factor $\tau_{\ell}(\eta)$ is suppressed by the lattice length $L$,
\begin{align}
\tau_{\ell>0}(\eta)
= 
\mathcal{O}\left(L^{-1}\right)\,.
\end{align}
In other words, the topological volume correction for $\ell > 0$ is smaller by a factor of $L$ relative to the $\ell = 0$ correction.  In our analysis of fermion-dimer scattering we have therefore included topological volume corrections as written in Eq.~(\ref{eqn:topologicalphases-029}) for $\ell = 0,$ but neglected the corrections for $\ell > 0$.  We find that this prescription gives good agreement with the continuum infinite-volume STM\ results for partial waves $\ell=0,1,2$.

\newpage

\bibliographystyle{apsrev}
\bibliography{MCadiabatic}

\end{document}